\documentclass[10pt, conference]{IEEEtran}

\usepackage{amsmath, graphicx, amssymb, subfigure, setspace, amsthm, color, bbm, enumerate}

\newtheorem{dfn}{Definition}

\newtheorem{theorem}{Theorem}

\newtheorem{lemma}{Lemma}

\newcommand{\sink}{\mathcal{R}}
\newcommand{\source}{\mathcal{S}}
\newcommand{\network}{\mathcal{G}}
\newcommand{\edges}{\mathcal{E}}
\newcommand{\nodes}{\mathcal{N}}

\newcommand{\F}{\mathbb{F}}

\newcommand{\stexp}{\mbox{$\mathbb{E}$}}    
\newcommand{\Prob}{\ensuremath{\mathbb{P}}}
   
\newcommand{\bigO}{\mbox{$\mathrm{O}$}}

\newcommand{\clear}[1]{\textbf{Clear:}}

\newcommand{\beq}{\begin{equation}}
\newcommand{\eeq}{\end{equation}}
\newcommand{\bea}{\begin{eqnarray}}
\newcommand{\eea}{\end{eqnarray}}

\def\qed{\quad \vrule height6.5pt width6pt depth0pt} 

\def\qed{\quad \vrule height6.5pt width6pt depth0pt} 
\long\def\symbolfootnote[#1]#2{\begingroup \def\thefootnote{\fnsymbol{footnote}}\footnote[#1]{#2}\endgroup}  

\begin{document}
\onecolumn
\doublespace


\title{Low-Complexity Near-Optimal Codes\\ for Gaussian Relay Networks}

\author{
\authorblockA{$\text{Theodoros K. Dikaliotis}^1$, $\text{Hongyi Yao}^1$, $\text{A. Salman Avestimehr}^2$, $\text{Sidharth Jaggi}^3$, $\text{Tracey Ho}^1$\\
$\hspace{1mm}^1$California Institute of
Technology $\hspace{1mm}^2$Cornell University $\hspace{1mm}^3$Chinese University of Hong Kong\\
{$^1$\{tdikal, tho\}@caltech.edu $\hspace{1mm}^1$yaohongyi03@gmail.com
$\hspace{1mm}^2$avestimehr@ece.cornell.edu $\hspace{1mm}^3$jaggi@ie.cuhk.edu.hk}}}

\maketitle

\begin{abstract}
We consider the problem of information flow over Gaussian relay networks. Similar to the recent work by Avestimehr \emph{et al.}~\cite{salman}, we propose network codes that achieve up to a constant gap from the capacity of such networks. However, our proposed codes are also computationally tractable. Our main technique is to use the codes of Avestimehr \emph{et al.} as inner codes in a concatenated coding scheme.
\end{abstract}

\section{Introduction}
\label{sec:Intro}

The recent work of~\cite{salman} parallels the classical network coding results~\cite{ahlswede00network} for wireless networks. That is, it introduces a quantize-map-and-forward scheme that achieves all rates up to a constant gap to the capacity of Gaussian relay networks, where this constant depends only on the network size, and varies linearly with it.  
However the computational complexity of encoding and decoding for the codes of~\cite{salman} grows exponentially in the block-length.

In this work, we aim to construct low complexity  coding schemes that achieve to within a constant gap of the capacity of Gaussian relay networks. The simplest Gaussian network that has been well investigated in the coding literature is the point-to-point Additive White Gaussian Noise (AWGN) channel. There have been a variety of capacity achieving codes developed for such channels (see {\it e.g.},~\cite{awgn_lattice, barron_codes, low_density_lattice_codes, multilevel_codes} and references therein). 

There have also been several recent efforts (for instance~\cite{sae_young_codes} and~ \cite{lattice_diggavi}) to extend the result of~ \cite{salman} and build \emph{low complexity} relaying strategies that achieve up to a constant gap from the capacity of Gaussian relay networks. However the decoding complexity of the proposed codes is still exponential in the block-length.
This is because these strategies are based on lattice codes, in which decoding proceeds via nearest neighbour search, which in general is computationally intractable.

In this paper, we build a coding scheme that has computational complexity that is polynomial in the block-length\footnote{The computational complexity of all existing codes, such as in~\cite{salman}, also grow exponentially in network parameters, namely the number of nodes $|\nodes|$. In fact, the codes proposed in this work have the same property. However, for our purposes we consider the network size to be fixed and small.} and achieves rates up to a constant gap to the capacity of Gaussian relay networks.

More specifically, we build a Forney-type~\cite{forney} two-layered concatenated code construction for Gaussian relay networks.
As our inner codes we use improved versions of the inner code in~ \cite{salman}, which can be decoded with probability of error dropping exponentially fast in the inner code block-length. As our outer codes we use polar codes~\cite{polar_codes}, which are provably capacity achieving (asymptotically in their block-lengths) for the Binary Symmetric Channel, and have computational complexity that is near-linear in their block-lengths.

\section{Model}
\label{sec:the_model}

We consider an additive white Gaussian noise (AWGN) relay network $ \network=\left(\nodes,\edges\right)$, where $\nodes$ denotes the set of nodes and $\edges$ the set of edges between nodes. Within the network there is a source $\source\in\nodes$ and a sink node\footnote{Just as in wired network coding, this result extends directly to the case of multiple sinks -- to ease notational burden we focus on the case of a single sink.}  $\sink\in\nodes$ where the source has a set of messages it tries to convey to the sink. For every node $i\in\nodes$ in the network there is the set $\mathcal{I}_i=\{j\in \nodes:(j,i)\in\edges\}\subseteq\nodes$ of nodes that have edges incoming to node $i$. All nodes have a single receiving and transmitting antenna and the received signal $y_{i,t}$ at node $i$ at time $t$ is given by
\begin{align} 
y_{i,t}=\sum_{j\in\mathcal{I}_i}h_{ji}x_{j,t}+z_{i,t}
\label{eqn:non_quantized_noise}
\end{align}
where $x_{j,t}\in \mathbb{C}$ the signal transmitted from node $j$ at time $t$ and $h_{ji}\in \mathbb{C}$ is the channel gain associated with the edge connecting nodes $j$ and $i$. The receiver noise, $z_{i,t}$, is modeled as a complex Gaussian random variable $ \mathcal{CN}(0,1)$, with i.i.d. distribution across time. Further we assume that there is an average transmit power constraint equal to $1$ at each node in the network.

Without loss of generality we assume that network $\network$ has a layered structure, \emph{i.e.}, all paths from the source to the destination have equal length. The number of layers in $\network$ is denoted by $L_\network$, and the source $\source$ resides at layer $1$ whereas the sink $\sink$ is at layer $L_\network$. As in  Section VI-B of ~\cite{salman} the results of this paper can be extended to the general case by expanding the network over time since the time-expanded network is layered.

\section{Transmission Strategy and Main Result} \label{sec:Coding_and_Relaying_Strategy}

\begin{figure}[ht]
\begin{center}
\includegraphics[width=1.0\columnwidth]{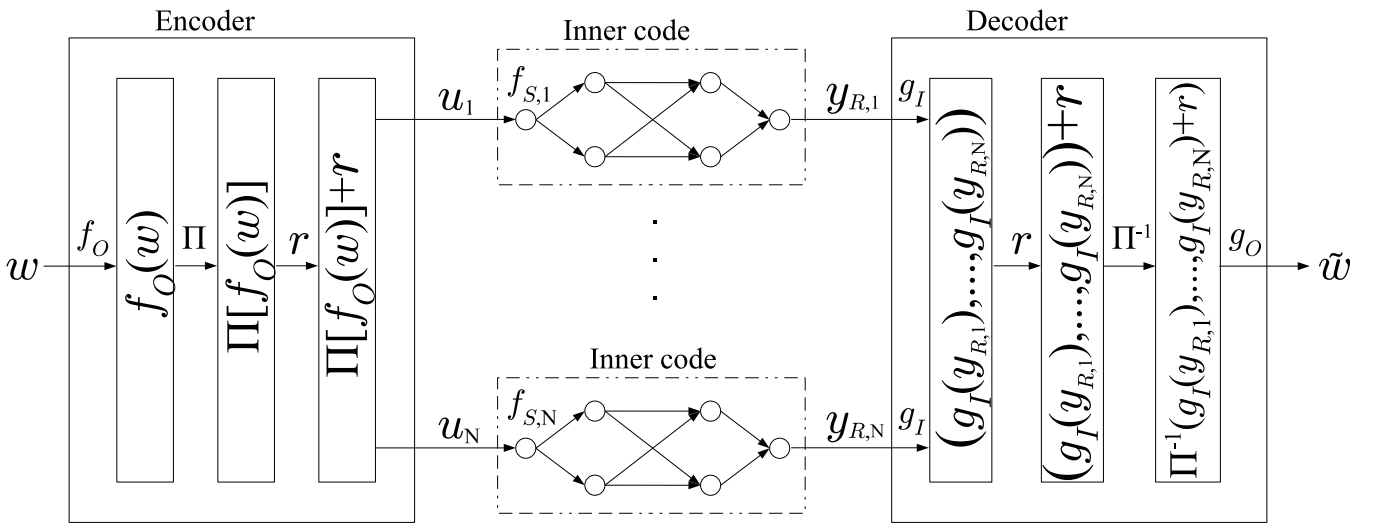}
\end{center}
\caption{System diagram of our concatenated code design.} \label{fig:system_diagram} 
\end{figure}

Our source encoder operates at two levels, respectively the outer code level and the inner code level. The outer code we use is a polar code~\cite{arikan09} (along with a random permutation of bits passed from the outer code to the encoders for the inner code -- this permutation is used for technical reasons that shall be described later). The inner code is a random code similar to the inner code used in~\cite{salman}. The relay nodes use ``quantize-map-and-forward'' as in~\cite{salman} over a short block-length ({\it i.e.}, the block-length of the inner code). Finally the receiver decodes the inner coding operations, inverts the random permutation inserted by the source encoder, and finally decodes the corresponding outer polar code.

The reason for using a two-layered concatenated code is to achieve both low encoding and decoding complexity, and a decoding error probability that decays almost exponentially fast with the block-length. In particular, as we describe in Section~ \ref{sec:decoding_of_the_inner_code}, our inner codes have a decoding algorithm that is based on exhaustive search. To ensure our inner codes' decoding complexity is tractable we set the block-length of the inner codes to be a ``relatively small'' fixed value. As a result, the decoding error probability for our inner codes is also fixed. To circumvent this, we then add an outer code on top of the inner code. More specifically, we use a polar code as an outer code, since polar codes have many desirable properties -- in particular, they provably have encoding and decoding complexities that are near linear in the block-length, are asymptotically capacity achieving, and also have probability of error that decays nearly exponentially in the block-length~ \cite{arikan09}.

One technical challenge arises. Polar codes are capable of correcting independent bit flips, but are not guaranteed to work against bursts of consecutive bit flips. It is to ``spread out" the possibly correlated bit-flips that would occur if one or several inner codes decode incorrectly that we introduce the random permutation mentioned above to the output of the source encoder's outer polar code. This random permutation is based on public randomness, and hence is available to both the source encoder, and the sink decoder (which can therefore invert it). The system diagram of our coding scheme is shown in Figure~ \ref{fig:system_diagram}. We now describe the details of the encoding, relaying, and decoding operations used by the source, relay nodes and the sink. Also, the source outer code encoder XORs a random string (denoted $r$, and known in advance to all parties via public randomness) to its output after permuting the bits of the polar code, but before passing these bits to the inner codes. This is to ensure independence of the inputs to the inner codes, so that concentration results can be used.

\subsection{Encoding at the Source and the Relays} \label{subsec:Encoding}

The overall communication scheme is over $n$ time instants and has a rate of $R=R_OR_I$, where $R_O$ and $R_I$ are the rates for the outer code, and each of the inner codes, respectively. The source takes a message $\mathbf{w}\in \F_2\hspace{-1mm}^{R_OR_In}$ of size $R_OR_In$ bits, and first applies the encoding algorithm of the polar code $f_O: 
\F_2\hspace{-1mm}^{R_OR_In}\rightarrow\F_2\hspace{-1mm}^{R_In}$
creating a string $f_O(\mathbf{w})\in\F_2\hspace{-1mm}^{R_In}$ of $R_In $ bits. Once the vector $f_O(\mathbf{w})$ is formed a random permutation $ \Pi$ is applied on all the bits of $f_O(\mathbf{w})$ giving rise to $ \Pi\left[f_O(\mathbf{w})\right]\in\F_2\hspace{-1mm}^{R_In}$. A random length-$R_In$ bit-vector $r$ is then XORed with $ \Pi\left[f_O(\mathbf{w})\right]\in\F_2\hspace{-1mm}^{R_In}$.

The derived bit-vector $\Pi\left[f_O(\mathbf{w})\right]$ is divided in $N=\frac{n}{\ell}$, \emph{i.e.} $\Pi\left[f_O(\mathbf{w})\right]= \left(\mathbf{u}_1,\ldots,\mathbf{u}_N\right)$,  chunks of size $R_I \ell$ bits, \emph{i.e.} $\mathbf{u}_k\in\F_2\hspace{-1mm}^{R_I\ell}$
for $1\leq k\leq N$. Each one of the bit-vectors $\mathbf{u}_k$ is conveyed through the network independently using a separate random inner code of rate $R_I$ and block-length $\ell$ introduced in~\cite{salman}.

More specifically, all nodes in the network operate over blocks of length $\ell$. The source randomly maps each inner code symbol $ \mathbf{u}_k\in\{1,\ldots,2^{R_I\ell}\}$ to a transmitted vector of length $\ell$ with components distributed i.i.d. from $\mathcal{CN} (0,1)$. That is, for each $1\leq k\leq N$ the random inner codes operate as $F_{\source,k}:\left\{1,2,\ldots,2^{R_I\ell}\right\}
\rightarrow \mathbb{C}\hspace{0.5mm}^\ell$, where realizations of $F_{\source,k}$ are denoted $f_{\source,k}$. For each $1\leq k\leq N$, each relay node $i\in\network$ rounds the real and imaginary part of each component of its received vector $\mathbf{y}_{i,k}\in\mathbb{C}\hspace{0.5mm}^\ell$ to the closest integer to form the length-$\ell$ vector ($[\mathbf{y}_{i,k}]$). It then randomly maps $ [\mathbf{y}_{i,k}]$ it to a transmit vector of length $\ell$ with components distributed i.i.d. from $\mathcal{CN}(0,1)$. For each $1\leq k\leq N$ the mappings at relay $i\in\nodes$ are denoted by $F_{i,k:}\left(\mathcal{A}_i^\ell, \mathcal{A}_i^\ell\right)\rightarrow\mathbb{C}\hspace{0.5mm}^\ell$,
with $f_{i,k}$ denoting a specific realization of $F_{i,k}$. Here $\mathcal{A}_i= \{-s_i,\ldots,s_i\}$ is the set of integers from $-s_i$ to $s_i$, where $s_i$ is a code design parameter be specified later. If the incoming signal at a relay node has a component such that its real or imaginary part is larger than $s_i$ in magnitude then an error is declared. Finally, the sink $\sink$ receives $\mathbf{y}_{\sink,k}\in\mathbb{C}\hspace{0.5mm}^\ell$
associated with inner code symbol  $\mathbf{u}_k$.

\subsection{Decoding}
\label{subsec:Decoding}

At the end of the transmission scheme, sink $\sink$ receives $ \mathbf{y}_\sink\in\mathbb{C}\hspace{0.5mm}^n$, which consists of $N$ chunks of length $\ell$, \emph{i.e.} $\mathbf{y}_\sink= \left(\mathbf{y}_{\sink,1},\ldots,\mathbf{y}_{\sink, N}\right)$ where $ \mathbf{y}_{\sink,k}\in\mathbb{C}\hspace{0.5mm}^\ell$. For each chunk $\mathbf{y}_{i,k}$, $1\leq k\leq N$, the sink $\sink$ then applies the decoding algorithm $g_I: 
\mathbb{C}\hspace{0.5mm}^\ell\rightarrow 2^{R_I\ell}$ of the corresponding inner code (to be described in Section~ \ref{sec:decoding_of_the_inner_code}). The sink then obtains $\hat{\mathbf{u}}_k$, a possibly noisy version of the corresponding inner code length-$\ell$ bit-vector $ \mathbf{u}_k$ that was transmitted during the $k^\text{th}$ application of the inner code. After decoding all chunks, it then XORs out the random bit-string $r$ added as part of the encoding procedure and then the inverse of the permutation applied by the source encoder is applied, \emph{i.e.} $ \Pi^{-1}\left[\left(g_I(\mathbf{y}_{\sink, 1}), \ldots,g_I(\mathbf{y}_{\sink, N})\right)\right]$. Finally, the sink utilizes the polar code decoder in order to produce the estimate for the source message, \emph{i.e.} $\tilde{\mathbf {w}}
=g_O \left(\Pi^{-1}\left[\left(g_I(\mathbf{y}_{\sink,1}),
\ldots,g_I(\mathbf{y}_{\sink,N})\right)\right]\right)$. An error is declared if ${\mathbf {w}} \neq \tilde{\mathbf {w}}$.

\subsection{Main Result}
\label{subsec:mainResult}

As in~\cite{salman} we define $\bar{C}$ to be the cut-set upper bound on the capacity $C$ of a general Gaussian relay network, {\it i.e.}, 
\begin{align*} C\leq \bar{C}\equiv\displaystyle\mathop{\max}_{p\left(\left\{x_j\right
\}_{j\in\nodes}\right)} \mathop{\min}_{\Omega\in\Lambda_\network} I \left(Y_{\Omega^c};X_\Omega\left|\right. X_{\Omega^c}\right)
\end{align*}
where $\Lambda_\network=\left\{\Omega:\source\in\Omega, \sink\in \Omega^c\right\}$ denotes the set of all source-sink cuts. We now state our main result.
\begin{theorem}
For any Gaussian relay network, the coding strategy described above achieves all rates within $(16|\nodes|+2)$ bits from the cut-set upper bound $\bar{C}$. This code has encoding complexity of $\bigO(n\log n+n 2^{\bar{C}\log\bar{C}})$,  decoding complexity of $\bigO(n\log n+n2^{(\bar{C}+13|\nodes|)\log_2\bar{C}})$, and a probability of error decaying as $\bigO(2^{-n^{-1/4}})$.
\label{thm:main_theorem}
\end{theorem}

The rest of the paper is devoted to proving this Theorem. First we give the decoding algorithm of the inner code that is based on exhaustively searching ``all possible'' noise patterns in the network. Then we continue specifying desirable parameter for our code.   

\section{Decoding of the inner code}
\label{sec:decoding_of_the_inner_code}

In~\cite{salman} they proved that by using their inner code--that is very similar to our inner code--the mutual information between the source message and what sink $\sink$ receives is within a constant gap from the capacity. Therefore by applying an outer channel code, any rate up to the mutual information and consequently rates within a constant gap from the capacity are achievable. In this work, despite our inner code is very similar to that in~\cite{salman} we do not use a mutual information type of argument but we instead devise an exhaustively search type of a decoding algorithm for the inner code. 

More specifically once the sink $\sink$ gets the received $\ell$--tuple vector $\left(y_1,\ldots,y_\ell\right)$ it rounds every component, \emph{i.e.} $\left([y_1],\ldots,[y_\ell]\right)$ and then it exhaustively tries all possible messages $u\in\{1,\ldots,2^{R_I\ell}\}$ and all highly probable noise patterns that give distinct outputs to find which was the source message that could have created the received signal. A decoding error is declared if there are more than one source messages $u$ that could have given the received signal or if there are no messages at all.

Since the noise at each node is a continuous random variable there is an infinite number of possible noises that can occur even at a single time instance. Therefore it is impossible to exhaustively search all possible noises that happened during all $\ell$ time steps of the inner code block-length. Due to the fact though that every relay node rounds its received signal to the closest integer only a countable number of noise patterns would give different received signals. Moreover we will show below that out of these infinite but countable many noises there is a finite set of noises that happen with high probability and the probability that there a noise pattern happens and does not belong to that set is very small.

\subsection{Quantized noise}
\label{subsec:quantised_noise}

In the following we will formalize all the above and we will start by giving the definition of quantized noise.
\begin{dfn}
The quantized noise $q_{i,t}$ for node $i\in\mathcal{N}$ at time $t$ is defined as
\begin{align}
q_{i,t}=\left[y_{i,t}\right]-\left[\sum_{j\in\mathcal{G}_i}h_{ji}x_{j,t}\right]
\label{eqn:quantized_noise}
\end{align}
where $[w]$ is the rounding of the real and imaginary part of the complex number $w\in\mathbb{C}$\hspace{0.8mm} to the closest integer.
\label{dfn:quatized_noise}
\end{dfn}

Since $y_{i,t}=\sum_{j\in\mathcal{G}_i}h_{ji}x_{j,t}+z_{i,t}$ it is shown in Appendix~\ref{appen:rounding} that the quantized noise $q_{i,t}$ can be written in the form
\begin{align}
q_{i,t}=\left[z_{i,t}\right]+R_{i,t}
\label{eqn:different_def_of_quantized_noise}
\end{align}
where $R_{i,t}$ can take any of the $9$ values $R_{i,t}=a_{i,t}+b_{i,t}\mathbbm{i}$ with $\mathbbm{i}^2=-1$, $a_{i,t}, b_{i,t}\in\{-1,0,1\}$ and $z_{i,t}$ is distributed as $\mathcal{CN}(0,1)$ for all $t\in\{1,\ldots,\ell\}$. Equation (\ref{eqn:quantized_noise}) gives that the quantized version of the received signal is given by the quantization of the transmitted signal plus the quantized noise. Therefore out of all the countably many quantized noise $\ell$-tuples $(q_{i,1},\ldots,q_{i,\ell})$ we want to find a finite set $Q_\ell$ called the ``candidate quantized noise set'' where the most probable quantized noise $\ell$--tuples are contained. We will use this set $Q_\ell$ for our exhaustive search algorithm.

\subsection{Candidate quantized noise set}
\label{subsec:candidate_quantized_noise_set}

One could possible choose the candidate quantized noise set $Q_\ell$ to be the typical set for random variable $q_{i,t}$ as defined in Chapter $3$ of~\cite{cover:bk}. The difficulty with this approach is that we do not know the distribution of $q_{i,t}$, since the distribution of $R_{i,t}$ is unknown and moreover random variables $z_{i,t}$ and $R_{i,t}$ are correlated. On the other hand we know the distribution of random variables $z_{i,t}$ and we will define set $Z_\ell$ to be:
\begin{dfn}
The set of $Z_\ell$ is the set of those $\ell$--tuples $\left([z_1],\ldots,[z_\ell]\right)\in\mathbb{Z}^\ell$ such that
\begin{align*}
p\left([z_1],\ldots, [z_\ell]\right)\geq 2^{-9\ell},
\end{align*}
where $p\left([z_1],\ldots, [z_\ell]\right)$ is the probability of the $\ell$--tuple and $z_i$ are i.d.d. $\mathcal{CN}(0,1)$ random variables.
\label{dfn:typical_quantized_normal_noise}
\end{dfn}
From the definition above\footnote{The entropy of the random variable $[z]$ where $z$ is distributed as $\mathcal{CN}(0,1)$ is around $4.4$ and therefore the typical set as defined in~\cite{cover:bk} will contain all $\ell$--tuples having probability $2^{-(4.4\pm\epsilon)\ell}$. Therefore set $Z_\ell$ contains the typical set for $\epsilon\leq 4.6$.} it is clear that set $Z_\ell$ has at most $2^{9\ell}$ elements, \emph{i.e.} $\left|Z_\ell\right|\leq 2^{9\ell}$. In Appendix~\ref{appen:outside_A_ell} it is proved that the probability of some $\ell$--tuple $\left([z_1],\ldots,[z_\ell]\right)$ ($z_i$ are i.i.d. $\mathcal{CN}(0,1)$ random variables) drawn randomly from $\mathcal{CN}(0,1)$ to be outside of $Z_\ell$ drops as $2^{-2\ell}$. Moreover in Appendix~\ref{appen:find_set_A_ell} a random procedure that is based on the ``Coupon Collector Problem'' is proposed on how to find all the elements in $Z_\ell$ with probability of failure (missing some elements of $Z_\ell$) dropping as fast as $2^{-\ell}$. 

The random variable $R_{i,t}$ defined in equation (\ref{eqn:different_def_of_quantized_noise}) takes $9$ possible values. Therefore there are $9^\ell$ possible $\ell$--tuples $\left(R_{i,1},\ldots,R_{i,\ell}\right)$ and the set that contains all $9^\ell$ such $\ell$--tuples is called set $R_\ell$. Now we are ready to give the definition of the candidate quantized noise set $Q_\ell$:  
\begin{dfn}
The candidate quantized noise set $Q_\ell$ is the set of all $\ell$ tuples of quantized noise such that
\begin{align*}
Q_\ell=\left\{z_i+r_j:\forall z_i\in Z_\ell\text{ and }r_j\in R_\ell\right\}
\end{align*}
denoting the sum of all $Z_\ell$ $\ell$--tuples with all possible $R_\ell$ $\ell$--tuples.
\label{dfn:typical_quantized_noise}
\end{dfn}
From the definition above along with Appendix~\ref{appen:find_set_A_ell} since we considers all possible $\ell$--tuples of $R_{i,t}$, no matter what the exact distribution of $R_{i,t}$ is, the probability that a quantized noise $\ell$--tuple $\mathbf{Q}=\left(q_{i,1},\ldots,q_{i,\ell}\right)$ randomly chosen will be outside $Q_\ell$ will be given by 
\begin{align}
\Prob\left(\mathbf{Q}\notin Q_\ell\right)\leq 2^{-2\ell}.
\label{eqn:outside_typical_quantized_noise}
\end{align}
It is easy to see that the size of set $Q_\ell$ is upper bounded by $\left|Q_\ell\right|\leq 9^\ell\hspace{0.3mm}\left|Z_\ell\right|\leq 9^\ell\hspace{0.3mm} 2^{9\ell}<2^{13\ell}$. From now on we will assume that if a quantized $\ell$--tuple that ``occurs'' in any node in the network is outside $Q_\ell$ then this is declared as an error and due to (\ref{eqn:outside_typical_quantized_noise}) this error happens with probability less than $2^{-2\ell}$.

\subsection{Probability of indistinguishability}
\label{subsec:indistinguishability}

As we discussed before once the sink $\sink$ rounds its received signal $\ell$--tuple vector $\left([y_1],\ldots,[y_\ell]\right)$ it exhaustively tries all possible messages $u\in\{1,\ldots,2^{R_I\ell}\}$ and all quantized noise tuples in $Q_\ell$ to find which was the source message that could have created the received signal. An error in the decoding of the inner code occurs if a noise pattern outside of $Q_\ell$ occurs in some node $i\in\nodes$. Then the decoding of the inner code fails and this happens with probability at most $|\nodes|2^{-\ell}$ according to (\ref{eqn:outside_typical_quantized_noise}) and the union bound over all nodes. Moreover if the noise patterns in all nodes happen inside $Q_\ell$ then the inner code fails if there are more than one messages $u$ that would have given rise to the same received signal and in the following we will analyze this probability $\Prob(u\rightarrow u')$.

Assume that the source node $\source$ sends message $u$ and the quantized noise realization $q_\network$ under message $u$ at all the nodes in the network is $q_\network=\mathbf{a}$. Then $\Prob\left(u\rightarrow u'\right)$ is the probability that there is another message $u'$ $(u'\neq u)$ and some noise realization $q'_\network=\mathbf{b}$ under message $u'$ (the two noise realizations are not necessarily different) so that the sink $\sink$ cannot distinguish whether message $u$ or $u'$ was sent. Similar to~\cite{salman} equation $(70)$ we have
\begin{align}
\hspace{-5mm}\Prob\left(u\rightarrow u'\left|q_\network=\mathbf{a}, q_\network=\mathbf{b}\right.\right)=\sum_{\Omega\in\Lambda_\network} \mathcal{P}_{\Omega,\mathbf{a},\mathbf{b}}
\label{eqn:general_indistinguishability_1}
\end{align}
where $\Omega$ is any cut in the network and $\mathcal{P}_{\Omega,\mathbf{a},\mathbf{b}}$ is defined to be the probability that nodes in $\Omega$ can distinguish between source messages $u$ and $u'$ under the noise realization $\mathbf{a}$ and $\mathbf{b}$ respectively while nodes in $\Omega^c$ cannot distinguish $u$ and $u'$. It is proved in Appendix~\ref{appen:prob_of_indistinguishability} that this probability is upper bounded by 
\begin{align}
\mathcal{P}_{\Omega,\mathbf{a},\mathbf{b}}\leq 2^{-\ell(\bar{C}-3\left|\nodes\right|)}.
\label{eqn:P_Omega_a_b_new}
\end{align}

Assume that the source node $\mathcal{S}$ sends message $u$ and the quantized noise realization in every node in the network is $\mathbf{a}$, then the probability that there is another message $u'$ or quantized noise $\ell$--tuple in $Q_\ell$ that will confuse the receiver is given by
\begin{align*}
\Prob(u\rightarrow &u')\leq \mathcal{P}_{\Omega, \mathbf{a}, \mathbf{b}}\hspace{0.3mm} 2^{R_I\ell}\hspace{0.3mm} \left|Q_\ell\right|^{\left|\nodes\right|}\displaystyle\mathop{\leq}^{(\ref{eqn:P_Omega_a_b_new})} 2^{-\ell\left(\bar{C}-3\left|\nodes\right|\right)} 2^{R_I\ell}\hspace{0.3mm} \left|Q_\ell\right|^{\left|\nodes\right|}\notag\\
&\leq
2^{-\ell\left(\bar{C}-3\left|\nodes\right|\right)} 2^{R_I\ell}\hspace{0.3mm} 2^{13\ell\left|\nodes\right|}\equiv 2^{-\ell\left(\bar{C}-16\left|\nodes\right|-R_I\right)}
\end{align*}
Therefore by setting the rate of the inner code
\begin{align}
R_I=\bar{C}-16\left|\nodes\right|-1
\label{eqn:equality_for_inner_rate}
\end{align} 
the probability that two messages $u$ and $u'$ will be indistinguishable at the receiver $\sink$ decays as 
\begin{align}
\Prob(u\rightarrow &u')\leq 2^{-\ell}
\label{eqn:overall_prob_of_indistinguishability}
\end{align}

Finally at each node we have some mappings $F_i:\left(\mathcal{A}_i^\ell, A_i^\ell\right)\rightarrow \mathbb{C}\hspace{0.5mm}^\ell$ where $\mathcal{A}_i$ is the set of integers $\left\{-s_i,\ldots,s_i\right\}$ so that the probability of a incoming signal having a component outside of $\mathcal{A}_i$ to be very small. Specifically for every node $i\in\mathcal{N}$ there is a set of signals that give the maximum absolute value $M_i$ for the received real or imaginary part, \emph{i.e.}
\begin{align*}
M_i=\displaystyle\mathop{\max}_{j\in\mathcal{G}_i,t}\left\{ \left|\text{Re}\left[\sum_{j\in\mathcal{G}_i}h_{ji}x_{j,t}\right]\right|, \left|\text{Im}\left[\sum_{j\in\mathcal{G}_i}h_{ji}x_{j,t}\right]~ \right|\right\}
\end{align*}
then $s_i=M_i+\delta_i$ where $\delta_i>0$ corresponds to the smallest slack necessary to make sure that the received signal (transmitted signal $+$ noise) is less in absolute value than $s_i$ with probability at most $2^{-2\ell}$. If the received signal has a real or imaginary part that its absolute value exceeds $s_i$ then the noise should have a real or imaginary part with absolute value larger than $\delta$, \emph{i.e.}
\begin{align*}
\Prob\left(z\geq\delta_i\right)\mathop{\leq}^{(*)}\text{Exp}\left(-\frac{\delta_i^2}{2}\right)\leq 2^{-2\ell}\Rightarrow \delta_i=\lceil\sqrt{\ell\hspace{0.3mm}2\hspace{0.3mm}\ln 2}\rceil.
\end{align*}
where inequality $(*)$ is derived by inequality $(7.1.13)$ at page $298$ of~\cite{abramowitz_stegun}.

\subsection{Probability of error for the inner code}

Now we are ready to analyze the overall probability of error for the inner code. The inner code fails if one of the following four events happen:
\begin{enumerate}
\item Some node $i\in\nodes$ in the network received a signal that its component has a real or an imaginary part with absolute value larger than $s_i$. According to the analysis above this event happens with probability less than $2^{-2\ell}\hspace{0.3mm}\left|\mathcal{N}\right|\leq 2^{-\ell}$ for large enough $\ell$ $\left(\ell\geq \log_2\left(\left|\nodes\right|\right)\right)$.

\item The random procedure that finds all the elements of set $Q_\ell$ failed. According to Appendix~\ref{appen:outside_A_ell} this happens with probability less than $2^{-\ell}$.

\item The quantized noise in some node $i\in\mathcal{N}$ in the network is outside of set $Q_\ell$ and according to (\ref{eqn:outside_typical_quantized_noise}) this happens with probability less than $2^{-2\ell}\hspace{0.3mm}\left|\mathcal{N}\right|\leq 2^{-\ell}$ for large enough $\ell$ $\left(\ell\geq \log_2\left(\left|\nodes\right|\right)\right)$. 

\item If the exhaustive search decoding procedure fails because there are more than one quantized noises $\ell$--tuples or messages that give the same signal to the receiver and according to (\ref{eqn:overall_prob_of_indistinguishability}) this happens with probability less than $2^{-\ell}$
\end{enumerate}
and therefore the overall probability of error $P_I$ for the inner code is upper bounded by
\begin{align}
P_I\leq 4\hspace{0.5mm} 2^{-\ell}
\label{eqn:probability_of_error_inner_code}
\end{align}
for $\ell>\log_2\left(\left|\mathcal{N}\right|\right)$.

\subsection{Complexity of the inner code}

In order to implement the inner code one have to find set $Q_\ell$. This is done by finding set $Z_\ell$ and the random approach based on the ``Coupon's Collector Problem'' to create set $Z_\ell$ requires $\bigO(\ell 2^{9\ell})$ order of steps. The most ``expensive'' operation though for the inner codes is the exhaustive decoding that search over all elements in $Q_\ell$ ($Q_\ell\leq 2^{13\ell}$) over all nodes in the network $|\nodes|$ and over all messages and therefore incurring and overall complexity $2^{R_I\ell}2^{R_I+13|\nodes|\ell}=2^{(R_I+13|\nodes|)\ell}\approx 2^{(\bar{C}+13|\nodes|)\ell}$.

\section{Complexity and error analysis of the code}
\label{sec:complexity_error_analysis}

We now specify the value of the parameter $\ell$ (which was left open thus far). For reasons that shall be clear in equation (\ref{eqn:rate}) for the rest of the paper we choose $\ell$ so that  
\begin{align}
h(2P_I)\leq\frac{1}{\bar{C}}
\label{eqn:value_of_PI}
\end{align}
where $h(x)=-x\log_2x-(1-x)\log_2(1-x)$ is the entropy function. The exact value for the probability of error for the inner code is not known but we can be pessimistic and take its largest value $P_I\leq 4\hspace{0.5mm} 2^{-\ell}$ given by~\ref{eqn:probability_of_error_inner_code}. One possible value for $\ell$ that satisfies inequality $(\ref{eqn:value_of_PI})$ as it is proved in Appendix~\ref{appen:inner_code_blocklength} is $\ell=3+\lceil\log_2\bar{C}\rceil$ and that is the value we will set the block length of the inner code to get. 

\subsection{Probability of error}

For each inner code that decodes its chunk incorrectly, in the worst case there is a burst of $R_I\ell$ erroneous bits that are passed to the sink's the polar code decoder. The purpose of the outer polar code is to correct these bit flips. Out of the $N=\frac{n}{\ell}$ input length-$\ell$ bit-vectors to the outer code on average only $\frac{n}{\ell}P_I$ are decoded erroneously (again, this number is concentrated around the expected value with high probability). This corresponds to $\frac{n}{\ell}P_IR_I\ell=nP_IR_I$ bit flips. Since each inner code chunk has independent inputs (due to the XORing operation described in the encoder) and is decoded independently we can apply the Chernoff bound and prove (Appendix~\ref{appen:chernoff1}) that the probability of having more than twice the expected number of bit flips drops at least as $\text{Exp}(-0.15\frac{n}{\bar{C}\log_2 \bar{C}})$.

The rate of the polar code is chosen $R_O=1-h(2P_I)$ can correct bit flips that are injected in our channel with probability only $P_I$. Therefore when less than twice the expected number of bit flips happen then the polar code fails with probability $2^{-(nR_I)^\beta}$ for any $\beta<1/2$ as proved in~\cite{arikan09} (the block-length of the polar code is $nR_I$). Therefore the probability of error of the overall code is dominated by the probability of error for the polar code is of the order $\bigO(2^{-n^{1/4}})$ for $\beta=1/4$.

\subsection{Achievable rate}

The achievable rate is $R=R_OR_I$ or
\begin{align}
R\geq \bar{C}-16|\nodes|-2
\label{eqn:rate}
\end{align}
since $R_I=\bar{C}-16|\nodes|-1$ and $R_O\geq 1-\frac{1}{\bar{C}}$ due to (\ref{eqn:value_of_PI}).

\subsection{Encoding decoding complexity}

The encoding and decoding complexity of our codes is the following:
\begin{itemize}

\item The encoding complexity for the outer polar code is $\bigO(n\log n )$ while the encoding complexity per inner code is $\bigO(2^{\bar{C}\log\bar{C}})$ so the overall encoding complexity is $\bigO(n\log n+n 2^{\bar{C}\log\bar{C}})$   

\item The decoding complexity of the polar code is $\bigO(n\log n)$ whereas the decoding complexity per inner code is $\bigO(2^{(\bar{C}+13|\nodes|)\log\bar{C}})$ so the overall encoding complexity is $\bigO(n\log n+n2^{(\bar{C}+13|\nodes|)\log_2\bar{C}})$.

\end{itemize}

\bibliographystyle{IEEEtran}
\bibliography{IEEEabrv,NWC-abbr}

\onecolumn
\allowdisplaybreaks

\begin{appendices}
\section{Proof of Lemma~\ref{lemma:prob_indistinguishability_mutual_information}}
\label{appen:prob_indistinguishability_mutual_information}

Consider the SVD decomposition of $\mathbf{H}$: $\mathbf{H}=\mathbf{U}\mathbf{\Sigma}\mathbf{V}^\dag$, with singular values $\sigma_1,\ldots,\sigma_{\min(n,m)}$. For every vector $\mathbf{f}\in\mathbb{C}\hspace{0.7mm}^{n\times 1}$ where $\left|\left|f\right|\right|_\infty\leq\sqrt{2}$ then $\left|\left|f\right|\right|_\infty\leq\sqrt{2n}$ and therefore
\begin{align}
\Prob\left(\forall 1\leq t\leq \ell: \left|\left|\mathbf{H}\left[\tilde{x}_{1,t},\ldots,\tilde{x}_{m,t}\right]^\text{T}+\mathbf{r}_t\right|\right|_\infty\leq\sqrt{2}\right)&\leq\Prob\left(\forall 1\leq t\leq \ell: \left|\left|\mathbf{H}\left[\tilde{x}_{1,t},\ldots,\tilde{x}_{m,t}\right]^\text{T}+\mathbf{r}_t\right|\right|_2\leq\sqrt{2n}\right)\notag\\
&=\Prob\left(\forall 1\leq t\leq \ell: \left|\left|\mathbf{U}\left(\mathbf{\Sigma}\mathbf{V}^\dag\left[\tilde{x}_{1,t},\ldots,\tilde{x}_{m,t}\right]^\text{T}+\mathbf{U}^\dag\mathbf{r}_t\right)\right|\right|_2\leq\sqrt{2n}\right)\notag\\
&\displaystyle\mathop{=}^{(a)}\Prob\left(\forall 1\leq t\leq \ell: \left|\left|\mathbf{\Sigma}\mathbf{V}^\dag\left[\tilde{x}_{1,t},\ldots,\tilde{x}_{m,t}\right]^\text{T}+\mathbf{U}^\dag\mathbf{r}_t\right|\right|_2\leq\sqrt{2n}\right)\notag\\
&\displaystyle\mathop{=}^{(b)}\Prob\left(\forall 1\leq j\leq \ell: \left|\left|\mathbf{\Sigma}\left[\tilde{x}_{1,t},\ldots,\tilde{x}_{m,t}\right]^\text{T}+\mathbf{\tilde{r}}_t\right|\right|_2\leq\sqrt{2n}\right)\notag\\
&=\Prob\left(\forall 1\leq t\leq \ell: \displaystyle\sum_{i=1}^{\min(n,m)}\left(\sigma_i\tilde{x}_{i,t}+\tilde{r}_{i,t}\right)^2+
\displaystyle\sum_{i=\min(n,m)+1}^n \tilde{r}_{i,t}^2\leq 2n\right)\notag\\
&=\Prob\left(\forall 1\leq t\leq \ell: \displaystyle\sum_{i=1}^{\min(n,m)}\left(\sigma_i\tilde{x}_{i,t}+\tilde{r}_{i,t}\right)^2 \leq 2n-\displaystyle\sum_{i=\min(n,m)+1}^n \tilde{r}_{i,t}^2\right)\notag\\
&=\prod_{t=1}^\ell\Prob\left(\displaystyle\sum_{i=1}^{\min(n,m)}\left(\sigma_i\tilde{x}_{i,t}+\tilde{r}_{i,t}\right)^2 \leq 2n-\displaystyle\sum_{i=\min(n,m)+1}^n \tilde{r}_{i,t}^2\right)
\label{eqn:appen:proof_of_lemma_prob_indistinguishability_mutual_information_eqn1}
\end{align}
where equality $(a)$ holds due to the fact that unitary matrices preserve the norm--$2$ of a vector, $(b)$ holds since $\mathbf{V}^\dag\left[\tilde{x}_{i,1},\ldots,\tilde{x}_{i,\ell}\right]^\text{T}$ have the same distribution with $\left[\tilde{x}_{i,1},\ldots,\tilde{x}_{i,\ell}\right]^\text{T}$ and $\mathbf{\tilde{r}}=\mathbf{U}^\dag\mathbf{r}$. We will denote by $P_j$ the following probability
\begin{align}
\mathcal{P}_j=\Prob\left(\displaystyle\sum_{i=1}^{\min(n,m)}\left(\sigma_i\tilde{x}_{i,t}+\tilde{r}_{i,t}\right)^2 \leq \omega\right)
\label{eqn:appen:proof_of_lemma_prob_indistinguishability_mutual_information_eqn2}
\end{align}
where $\omega=2n-\displaystyle\sum_{i=\min(n,m)+1}^n \tilde{r}_{i,t}^2$ and therefore 
\begin{align}
P_j&=\Prob\left(\displaystyle\sum_{i=1}^{\min(n,m)}\left(\sigma_i\tilde{x}_{i,t}+\tilde{r}_{i,t}\right)^2 \leq \omega\right)\notag\\
&=\Prob\left(-k\displaystyle\sum_{i=1}^{\min(n,m)}\left(\sigma_i\tilde{x}_{i,t}+\tilde{r}_{i,t}\right)^2 \geq -k\hspace{0.1mm}\omega\right)\ \ \forall k>0\notag\\
&=\Prob\left(\text{Exp}\left[-k\displaystyle\sum_{i=1}^{\min(n,m)}\left(\sigma_i\tilde{x}_{i,t}+\tilde{r}_{i,t}\right)^2\right] \geq\text{Exp}\left[-k\hspace{0.1mm}\omega\right]\right)\notag\\
&\displaystyle\mathop{\leq}^{(c)}\stexp\left\{\text{Exp}\left[-k\displaystyle\sum_{i=1}^{\min(n,m)}\left(\sigma_i\tilde{x}_{i,t}+\tilde{r}_{i,t}\right)^2\right]\right\} \text{Exp}\left[k\hspace{0.1mm}\omega\right]\notag\\
&\displaystyle\mathop{=}^{(d)}\text{Exp}\left[k\hspace{0.1mm}\omega\right]\prod_{i=1}^{\min(n,m)}\stexp\left\{\text{Exp}\left[-k\left(\sigma_i\tilde{x}_{i,t}+\tilde{r}_{i,t}\right)^2\right]\right\}\notag\\
&=\text{Exp}\left[k\hspace{0.1mm}\omega\right]\prod_{i=1}^{\min(n,m)}\frac{1}{\sqrt{2\pi\sigma^2}}\int_{-\infty}^{+\infty}\text{Exp}\left[-k(\sigma_i y+\tilde{r}_{i,t})^2\right] \text{Exp}\left[-\frac{y^2}{2\sigma^2}\right]dy\notag\\
&=\text{Exp}\left[k\hspace{0.1mm}\omega\right]\prod_{i=1}^{\min(n,m)}\frac{\text{Exp}\left(-\frac{\tilde{r}_{i,j}^2t}{1+2\sigma_i^2\sigma^2t}\right)}{\sqrt{1+2\sigma_i^2\sigma^2t}},
\label{eqn:appen:proof_of_lemma_prob_indistinguishability_mutual_information_eqn3}
\end{align}
where $(c)$ is the Markov inequality and $(d)$ holds since $\tilde{x}_{i,t}$ are independent for different $t$. From equation (\ref{eqn:appen:proof_of_lemma_prob_indistinguishability_mutual_information_eqn3}) for $k=\frac{1}{2\sigma^2}$ we get
\begin{align*}
P_j&\leq\text{Exp}\left[\frac{\omega}{2\sigma^2}\right]\prod_{i=1}^{\min(n,m)}\frac{\text{Exp}\left(-\frac{\tilde{r}_{i,t}^2}{2\sigma^2(1+\sigma_i^2)}\right)}{\sqrt{1+\sigma_i^2}}\\
&=\text{Exp}\left[\frac{1}{2\sigma^2}\left(\omega-\sum_{i=1}^{\min(n,m)}\frac{\tilde{r}_{i,t}^2}{1+\sigma_i^2}\right)\right] \text{Exp}\left[-\frac{1}{2}\sum_{i=1}^{\min(n,m)}\ln(1+\sigma_i^2)\right]
\end{align*}
and by substituting $\omega$ with its value $\omega=2n-\displaystyle\mathop{\sum}_{i=\min(n,m)+1}^{n}\tilde{r}_{i,t}^2$ along with the equation above we get
\begin{align*}
P_j&\leq\text{Exp}\left[\frac{1}{2\sigma^2}\left(2n-\sum_{i=\min(n,m)+1}^n\tilde{r}_{i,t}^2-\sum_{i=1}^{\min(n,m)}\frac{\tilde{r}_{i,t}^2}{1+\sigma_i^2}\right)\right] \text{Exp}\left[-\frac{1}{2}\sum_{i=1}^{\min(n,m)}\ln(1+\sigma_i^2)\right]\\
&\leq\text{Exp}\left[\frac{1}{2\sigma^2}\left(2n\right)\right] \text{Exp}\left[-\frac{1}{2}\sum_{i=1}^{\min(n,m)}\ln(1+\sigma_i^2)\right]\\
&\leq\text{Exp}\left[-\frac{1}{2}\left(\sum_{i=1}^{\min(n,m)}\ln(1+\sigma_i^2)-\frac{2n}{\sigma^2}\right)\right]
\end{align*}
and combining the inequality above with (\ref{eqn:appen:proof_of_lemma_prob_indistinguishability_mutual_information_eqn1}) we get
\begin{align*}
\Prob\left(\forall 1\leq j\leq \ell: \left|\left|\mathbf{H}\left[\tilde{x}_{1,j},\ldots,\tilde{x}_{m,j}\right]^\text{T}+\mathbf{r}_j\right|\right|_\infty\leq\sqrt{2}\right) &\leq\text{Exp}\left[-\frac{\ell}{2}\left(\sum_{i=1}^{\min(n,m)}\ln(1+\sigma_i^2)-\frac{2n}{\sigma^2}\right)\right]\\
&\leq 2^{-\left[\frac{\ell}{2}\left(\sum_{i=1}^{\min(n,m)}\ln(1+\sigma_i^2)-\frac{2n}{\sigma^2}\right)\right]\log_2e}\\
&\leq 2^{-\frac{\ell}{2}\left(\sum_{i=1}^{\min(n,m)}\log_2(1+\sigma_i^2)-\frac{2n}{\sigma^2}\log_2e\right)}
\end{align*}
and since, $\frac{1}{2}\sum_{i=1}^{\min(m,n)}\log_2(1+\sigma_i^2)=\frac{1}{2}\log_2\text{det} (\mathbf{I}_n+\mathbf{H}\mathbf{H}^\dagger)=I\left(\mathbf{x};\mathbf{H}\mathbf{x}+\mathbf{z}\right)$, for $\mathbf{x}$, $\mathbf{z}$ that are distributed as $\mathcal{CN}\left(\mathbf{0},\mathbf{I}_m\right)$ and $\mathcal{CN}\left(\mathbf{0},\mathbf{I}_n\right)$ respectrively we get
\begin{align}
\Prob\left(\forall 1\leq j\leq \ell: \left|\left|\mathbf{H}\left[\tilde{x}_{1,j},\ldots,\tilde{x}_{m,j}\right]^\text{T}+\mathbf{r}_j\right|\right|_\infty\leq\sqrt{2}\right) \leq 2^{-\ell\left(I\left(\mathbf{x};\mathbf{H}\mathbf{x}+\mathbf{z};\right)-\frac{n}{\sigma^2}\log_2e\right)}
\label{eqn:appen:proof_of_lemma_prob_indistinguishability_mutual_information_eqn4}
\end{align}

\section{Rounding to the closest integer the sum of two numbers}
\label{appen:rounding}

\begin{figure}[ht]
\centering
\subfigure[The value of $L$ for $\text{[}h_1-h_2\text{]}=\text{[}h_1\text{]}-\text{[}h_2\text{]}+L$]{
\includegraphics[scale=0.3]{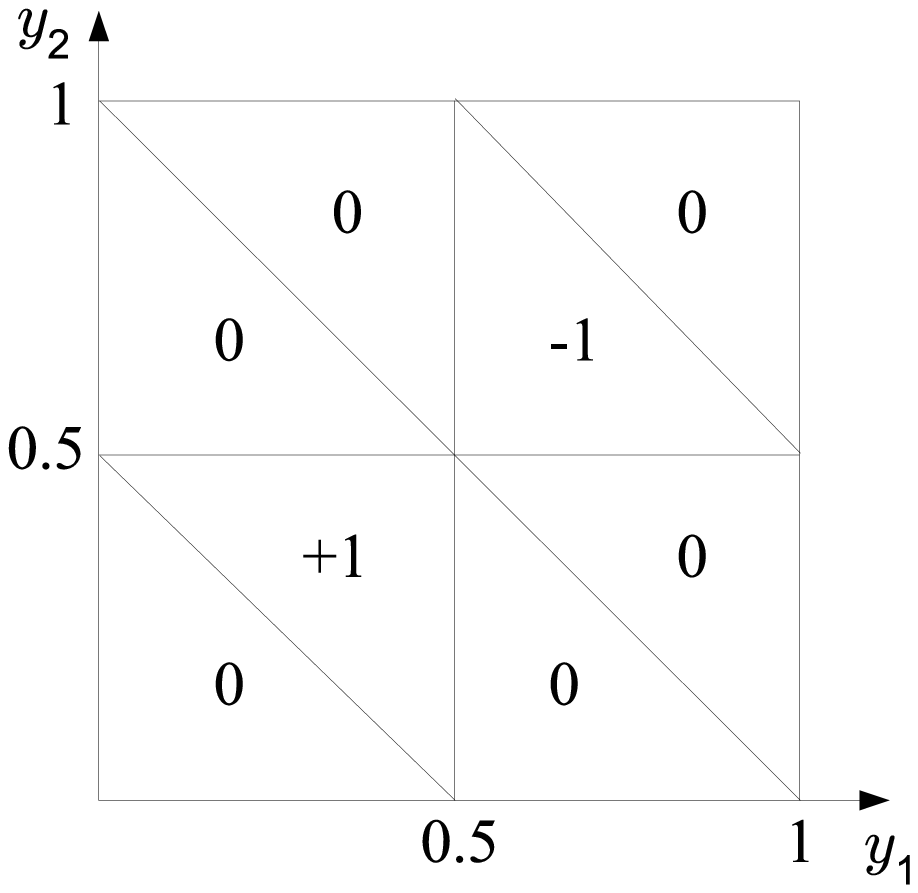}
\label{fig:value_of_L_positive}}
\subfigure[The value of $L$ for $\text{[}h_1-h_2\text{]}=\text{[}h_1\text{]}-\text{[}h_2\text{]}+L$]{
\includegraphics[scale=0.3]{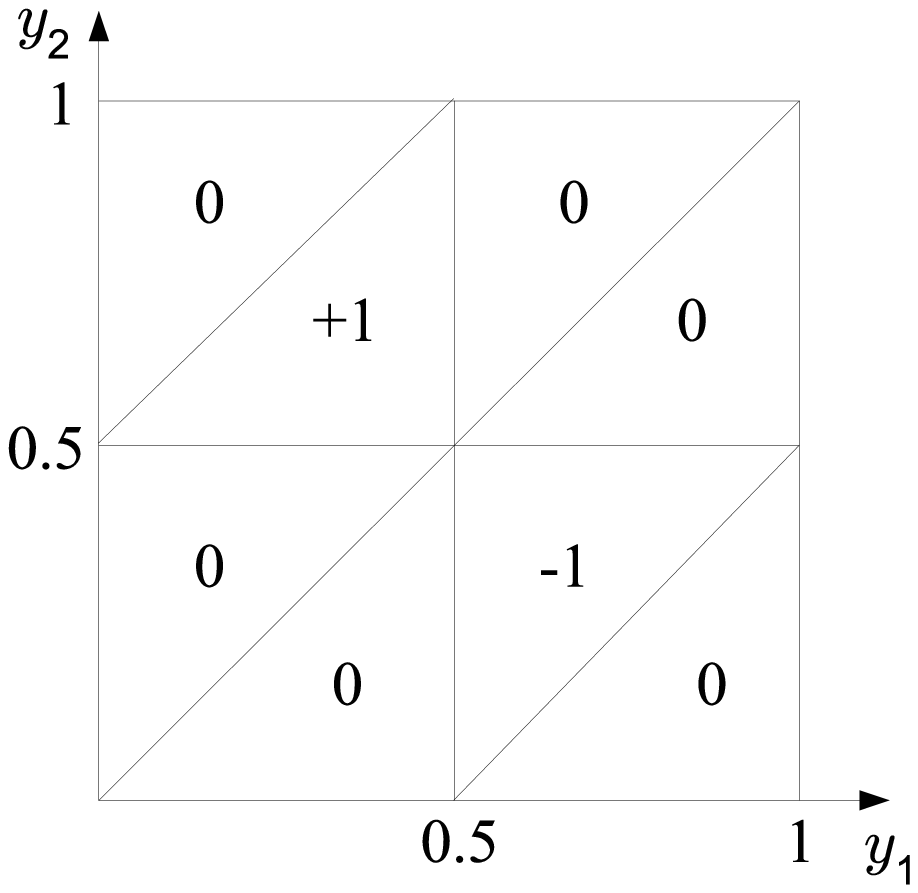}
\label{fig:value_of_L_negative}}
\caption{The value of $L$ for different regions of $(y_1,y_2)$ and different operations (addition \& subtraction)} 
\end{figure}

We define as $[x]$ to be the nearest integer to some real number $x$ and if $z=a+bi\in\mathbb{C}$ for $a,b\in\mathbb{R}$ and $i^2=-1$ then $[z]=[a]+[b]i$. Assume that $h_1=x_1+y_1$ and $h_2=x_2+y_2$ are two positive real numbers with $x_1$, $x_2$ and $y_1$, $y_2$ representing the integer and decimal part of these numbers. Then if we write 
\begin{align*}
[h_1+h_1]=[h_1]+[h_2]+L
\end{align*}
the different values of $L$ for different regions of $(y_1,y_2)$ are given in Figure~\ref{fig:value_of_L_positive}. Similarly if we take the difference of $h_1$ and $h_2$ and we write
\begin{align*}
[h_1+h_1]=[h_1]+[h_2]+L
\end{align*}
then the values of $L$ for different regions of $(y_1,y_2)$ are shown in Figure~\ref{fig:value_of_L_negative}.

Therefore in general for two complex numbers $z_1, z_2\in\mathbb{C}$ we have
\begin{align*}
[z_1+z_2]=[z_1]+[z_2]+R
\end{align*}
where $R$ can take one of the $9$ values $R=r_1+r_2i$ where $r_1,r_1\in\{-1,0,1\}$.

\section{A random approach to find all elements in set $Z_\ell$}
\label{appen:find_set_A_ell}

The way to find all the elements in set $Z_\ell=\{z_1,\ldots,z_{\left|Z_\ell\right|}\}$ is based on the well known ``Coupon Collector's Problem''~\cite{mitzenmacher05probability}. Let's assume that one creates $r$ (the value of $r$ will be specified later) $\ell$--tuples $([z_1],\ldots,[z_\ell])$ (where $z_i$ are i.i.d. random variables distributed as $\mathcal{CN}(0,1)$) and denote as $A^r_\ell$ the set of all these $r$ randomly created $\ell$--tuples.

The probability that element $z_i\in Z_\ell$ is not contained in the randomly created set $A^r_\ell$ is given by
\begin{align*}
\Prob\left(z_i\notin A^r_\ell\right)=\left(1-p_i\right)^r\leq e^{-rp_i}
\end{align*}
where $p_i$ is the probability of element $z_i\in Z_\ell$ and since $p_i\geq 2^{-9\ell}\equiv Q$ then
\begin{align*}
\Prob\left(z_i\notin A^r_\ell\right)\leq e^{-rQ}.
\end{align*}
Moreover from the definition of $Q$ we get that $|Z_\ell|\leq \frac{1}{Q}$ and in general the probability that there is an element in $Z_\ell$ that is not contained in $A^r_\ell$ is
\begin{align*}
\Prob\left(\displaystyle\mathop{\bigcup}_{z_i\in Z_\ell}\{ z_i\notin A^r_\ell\}\right)\leq \sum_{z_i\in Z_\ell} \Prob\left(z_i\notin A^r_\ell\right)\leq \left|Z_\ell\right| e^{-rQ}\equiv \frac{1}{Q}e^{-rQ}
\end{align*}
and therefore if $r=\frac{10}{9}\frac{1}{Q}\ln\left(\frac{1}{Q}\right)$ we have that 
\begin{align*}
\Prob\left(\displaystyle\mathop{\bigcup}_{z_i\in Z_\ell}\{ z_i\notin A^r_\ell\}\right)\leq \frac{1}{Q}e^{-\frac{10}{9}\frac{1}{Q}\ln\left(\frac{1}{Q}\right)Q}=Q^{\frac{1}{9}}
\end{align*}
and since $Q=2^{-9\ell}$
\begin{align*}
\Prob\left(\displaystyle\mathop{\bigcup}_{z_i\in Z_\ell}\{ z_i\notin A^r_\ell\}\right)\leq 2^{-\ell}
\end{align*}
and the number of elements in $A^r_\ell$ is $10\hspace{0.3mm} \ell\hspace{0.3mm} 2^{9\ell}\hspace{0.3mm}\ln 2$.

\section{The probability that a randomly picked $\ell$--tuple will be outside set $Z_\ell$}
\label{appen:outside_A_ell}	

Assume that there is an $\ell$--tuple $\mathbf{Z}=\left([z_1],\ldots,[z_\ell]\right)$ drawn randomly where $z$ are i.i.d. random variables from distribution $\mathcal{CN}(0,1)$ then the probability that does not belong to $Z_\ell$ is 
\begin{align}
\Prob\left[\mathbf{Z}\notin Z_\ell\right]=\Prob\left[\prod_{i=1}^\ell p([z_i])< 2^{-9\ell}\right] = \Prob\left[\prod_{i=1}^\ell \frac{1}{p([z_i])}> 2^{9\ell}\right] = \Prob\left[\prod_{i=1}^\ell \left(\frac{1}{p([z_i])}\right)^k> 2^{9k\ell}\right]\ \forall k>0\notag\\
\displaystyle\mathop{\leq}^{(a)}\frac{\stexp\left[\prod_{i=1}^\ell \left(\frac{1}{p([z_i])}\right)^k\right]}{2^{9k\ell}}
\displaystyle\mathop{\leq}^{(b)}2^{-9k\ell} \prod_{i=1}^\ell \stexp\left[ \left(\frac{1}{p([z_i])}\right)^k\right]
=2^{-9k\ell}\left(\stexp\left[ p^{-k}([z])\right]\right)^\ell
=\left(2^{-9k}\stexp\left[ p^{-k}([z])\right]\right)^\ell
\displaystyle\mathop{=}^{(c)} \left(2^{-9k}\stexp^2\left[ p^{-k}([z_R])\right]\right)^\ell
\label{eqn:appen:outside_A_ell_1}
\end{align}
where $(a)$ is due to Markov inequality and $(b)$ is due to the independence of the $p([z_i])$ for $i\in\{1,\ldots,\ell\}$. At equality $(c)$, $z_R$ is random variable distributed as $\mathcal{N}(0,1)$ and the equality holds since the real and the imaginary part of a $\mathcal{CN}(0,1)$ random variable are independent $\mathcal{N}(0,1)$ random variables. If we define $c_i$ where $i\in\mathbb{Z}$ to be
\begin{align*}
c_i=\frac{1}{\sqrt{2\pi}}\int_{i-\frac{1}{2}}^{i+\frac{1}{2}}e^{-\frac{x^2}{2}}dx
\end{align*}
then 
\begin{align*}
\stexp\left[p^k\left([z_R]\right)\right]=\sum_{i=-\infty}^{\infty}c_i^{1-k}=c_0^{1-k}+ 2\sum_{i=1}^{\infty}c_i^{1-k}.
\end{align*}

One can bound $c_i$ by inequality $(7.1.13)$ at page $298$ of~\cite{abramowitz_stegun}: 
\begin{align*}
c_k\leq \frac{\text{Exp}\left(-k^2/2\right)}{k}\left(\frac{1}{\sqrt{2\pi}} -\sqrt{\frac{2}{9\pi}}\right)\ \text{for }k>0
\end{align*}
and approximating $c_0\leq 0.5$. From the above inequalities one can find that for $k=0.6$, $\stexp\left[p^k\left([z_R]\right)\right]\leq 2.9$ and therefore from inequality (\ref{eqn:appen:outside_A_ell_1}) we get
\begin{align*}
\Prob\left[\mathbf{Z}\notin Z_\ell\right]\leq \frac{2.9^2}{2^{9\cdot 0.6}}\leq (0.2)^\ell\leq 2^{-2\ell}.
\end{align*}

\section{Analyzing probability $\mathcal{P}_{\Omega,\mathbf{a},\mathbf{b}}$}
\label{appen:prob_of_indistinguishability}

The analysis is very similar to~\cite{salman}. We define 
\begin{align}
\mathcal{P}_{\Omega,\mathbf{a},\mathbf{b}}=\Prob\left(\text{Nodes in $\Omega$ can distinguish between $u$, $u'$ and nodes in $\Omega^\text{c}$ cannot}\left|q_\network=\mathbf{a}, q'_\network=\mathbf{b}\right.\right)
\label{eqn:general_indistinguishability_2}
\end{align}
where $\Omega$ is any cut in the network. We define the following sets and events:
\begin{itemize}
\item $L_l(\Omega)$: The nodes that are in $\Omega$ and are at layer $l$.
\item $R_l(\Omega)$: The nodes that are in $\Omega^\text{c}$ and are at layer $l$.
\item $\mathcal{L}_l(\Omega)$: The event that the nodes in $L_l(\Omega)$ can distinguish between $u$ and $u'$. 
\item $\mathcal{R}_l(\Omega)$: The event that the nodes in $R_l(\Omega)$ can not distinguish between $u$ and $u'$. 
\end{itemize}

The nodes in any set $A$ cannot distinguish between the message $u$ and $u'$ if the integer values of their received signals are identical, \emph{i.e.} $[y_A(u)]=[y_A(u')]$. Assume that the network is layered and there are $L_\network$ layers in total (the source $\source$ is at layer $L=1$ and the receiver $\sink$ is at layer $L_\network$ ). Therefore equation (\ref{eqn:general_indistinguishability_2}) becomes 
\begin{align*}
\mathcal{P}_{\Omega,\mathbf{a},\mathbf{b}}&=\Prob\left(\mathcal{R}_l(\Omega), \mathcal{L}_{l-1}(\Omega),l=2,\ldots,l_D\left|q_\network=\mathbf{a}, q'_\network=\mathbf{b}\right.\right)\\
&=\prod_{l=2}^{L_\network}\Prob\left(\mathcal{R}_l(\Omega), \mathcal{L}_{l-1}(\Omega)\left|\mathcal{R}_j(\Omega), \mathcal{L}_{j-1}(\Omega),j=2,\ldots,l-1,q_\network=\mathbf{a}, q'_\network=\mathbf{b}\right.\right)\\
&\leq\prod_{l=2}^{L\network}\Prob\left(\mathcal{R}_l(\Omega)\left|\mathcal{R}_j(\Omega), \mathcal{L}_{j}(\Omega),j=2,\ldots,l-1,q_\network=\mathbf{a}, q'_\network=\mathbf{b}\right.\right)\\
&\displaystyle\mathop{=}^{(*)}\prod_{l=2}^{L_\network}\Prob\left(\mathcal{R}_l(\Omega)\left|\mathcal{R}_{l-1}(\Omega), \mathcal{L}_{l-1}(\Omega),q_\network=\mathbf{a}, q'_\network=\mathbf{b}\right.\right)\\
&=\prod_{l=2}^{L_\network}\Prob\left(\left[y_{R_l(\Omega)}(u)\right]=\left[y_{R_l(\Omega)}(u')\right] \left|\mathcal{R}_{l-1}(\Omega),\mathcal{L}_{l-1}(\Omega),q_\network=\mathbf{a}, q'_\network=\mathbf{b}\right.\right)\\
&=\prod_{l=2}^{L_\network}\Prob\left(\forall 1\leq t\leq \ell:\left[\mathbf{H}_l x_{L_{l-1}(\Omega),t}(u)\right]+\mathbf{a}_{R_l(\Omega),t}=\left[\mathbf{H}_l x_{L_{l-1}(\Omega),t}(u')\right]+\mathbf{b}_{R_l(\Omega),t} \left|\mathcal{R}_{l-1}(\Omega),\mathcal{L}_{l-1}(\Omega),q_\network=\mathbf{a}, q'_\network=\mathbf{b}\right.\right)\\
&\mathop{=}^{(**)}\prod_{l=2}^{L_\network}\Prob\left(\forall 1\leq t\leq \ell:\left[\mathbf{H}_l x_{L_{l-1}(\Omega),t}(u)+\mathbf{a}_{R_l(\Omega),t}\right]=\left[\mathbf{H}_l x_{L_{l-1}(\Omega),t}(u')+\mathbf{b}_{R_l(\Omega),t}\right] \left|\mathcal{R}_{l-1}(\Omega),\mathcal{L}_{l-1}(\Omega),q_\network=\mathbf{a}, q'_\network=\mathbf{b}\right.\right)
\end{align*}
where $\mathbf{H}_l\in\mathbb{C}\hspace{0.7mm}^{\left|R_{l}(\Omega)\right|\times \left|L_{l-1}(\Omega)\right|}$ is the transfer matrix from the nodes in $L_{l-1}(\Omega)$ to the nodes in $R_{l}(\Omega)$. Vectors $x_{L_{l-1}(\Omega),t}(u)$, $x_{L_{l-1}(\Omega),t}(u')$ are the signals transmitted at time step $t$ from nodes in $L_{l-1}(\Omega)$ when the source $\source$ has transmitted messages $u$ and $u'$ respectively whereas $\mathbf{a}_{R_l(\Omega),t}$, $\mathbf{b}_{R_l(\Omega),t}$ are the noise realizations for nodes in $R_l(\Omega)$ at time $t$. Inequality $(*)$ holds due to the Markov structure of the network and $(**)$ the last equality holds since the components of $\mathbf{a}$, $\mathbf{b}$ are integers. 

Note that if $\mathbf{A}, \mathbf{B}\in \mathbb{C}\hspace{0.7mm}^{m\times 1}$ are complex vectors then
\begin{align*}
\left[\mathbf{A}_i\right]=\left[\mathbf{B}_i\right]\ \forall i\Rightarrow \left|\left|\mathbf{A}-\mathbf{B}\right|\right|_\infty\leq \sqrt{2}
\end{align*} 
and therefore from the previous equation we get
\begin{align*}
\mathcal{P}_{\Omega,\mathbf{a},\mathbf{b}}\leq\prod_{l=2}^{L_\network}\Prob\left(\forall 1\leq t\leq \ell:\left|\left|\mathbf{H}_l\left(x_{L_{l-1}(\Omega),t}(u)-x_{L_{l-1}(\Omega),t}(u')\right)+ \left(\mathbf{a}-\mathbf{b}\right)_{R_l(\Omega),t}\right|\right|_\infty\leq\sqrt{2} \left|\mathcal{R}_{l-1}(\Omega),\mathcal{L}_{l-1}(\Omega),z_\mathcal{V}=\mathbf{a}, z'_\mathcal{V}=\mathbf{b}\right.\right).
\end{align*}

We are now extending Lemma $1$ of~\cite{salman} in the following Lemma that is proved in Appendix~\ref{appen:prob_indistinguishability_mutual_information}.
\begin{lemma}
Assume $\left[\tilde{x}_{i,1},\ldots,\tilde{x}_{i,\ell}\right]$ for $i=1,\ldots,m$ are vectors of length $\ell$ with elements chosen i.i.d. from $\mathcal{CN}(0,\sigma^2)$. Moreover $\mathbf{r}_j\in \mathbb{C}\hspace{0.7mm}^{n\times 1}$ are a collection of vectors and $\mathbf{H}\in\mathbb{C}\hspace{0.7mm}^{n\times m}$ is some matrix then
\begin{align*}
\Prob\left(\forall 1\leq j\leq \ell: \left|\left|\mathbf{H}\left[\tilde{x}_{1,j},\ldots,\tilde{x}_{m,j}\right]^\text{T}+\mathbf{r}_j\right|\right|_\infty\leq\sqrt{2}\right) \leq 2^{-\ell\left(I\left(\mathbf{x};\mathbf{H}\mathbf{x}+\mathbf{z}\right)-\frac{n}{\sigma^2}\log_2e\right)}
\end{align*}
where $\mathbf{x}$, $\mathbf{z}$ are distributed as $\mathcal{CN}\left(\mathbf{0},\mathbf{I}_m\right)$ and $\mathcal{CN}\left(\mathbf{0},\mathbf{I}_n\right)$ respectrively.
\label{lemma:prob_indistinguishability_mutual_information}
\end{lemma}

Therefore since $\left(x_{L_{l-1}(\Omega),t}(u)-x_{L_{l-1}(\Omega),t}(u')\right)$ are Gaussian random variables distributed i.i.d. from $\mathcal{CN}(0,2)$ we have from applying Lemma~\ref{lemma:prob_indistinguishability_mutual_information} that
\begin{align}
\mathcal{P}_{\Omega,\mathbf{a},\mathbf{b}}&\leq \prod_{l=2}^{l_D}2^{-\ell\left(I\left(\mathbf{x}_{L_{l-1}(\Omega)};\mathbf{y}_{R_l(\Omega)}\left|\right.\mathbf{x}_{R_{l-1}(\Omega)}\right) -\frac{\left|R_l(\Omega)\right|}{2}\log_2e\right)}\leq 2^{-\ell\left(\bar{C}_\text{iid}-\left|\mathcal{V}\right|\right)}
\label{eqn:P_Omega_a_b}
\end{align}
where $\bar{C}_\text{iid}$ is defined as
\begin{dfn}
We define 
\begin{align*}
\bar{C}_\text{iid}=\displaystyle\min_\Omega I\left(x_\Omega;y_{\Omega^c}\left|\right.x_{\Omega^c}\right)
\end{align*}
where $x_i$, $i\in\mathcal{V}$, are i.i.d. $\mathcal{CN}(0,1)$ random variables and $\Omega$ is any cut in the network.
\label{dfn:Min_cut} 
\end{dfn}

Finally if we combine (\ref{eqn:P_Omega_a_b}) with Lemma $6.6$ in~\cite{salman} that gives $\bar{C}-\bar{C}_{iid}<2|\nodes|$ we conclude that $\mathcal{P}_{\Omega,\mathbf{a},\mathbf{b}}\leq 2^{-\ell(\bar{C}-3\left|\nodes\right|)}$.

\section{Evaluation of block-length of the inner code}
\label{appen:inner_code_blocklength}

We want to find a probability a value for $x\leq\frac{1}{2}$ such that $h(x)\leq \frac{1}{\bar{C}}$ where $h(x)=-x\log_2 x-(1-x)\log_2(1-x)$ is the entropy function. It requires simple algebra to prove that $h(x)< -2x\log_2 x$ for $x<0.4$. Therefore if $-2x\log_2 x\leq \frac{1}{\bar{C}}$ then $h(x)<\frac{1}{\bar{C}}$ for $x<0.4$. For $x=\frac{1}{\bar{C}^2}$ we get $-\frac{2}{\bar{C}^2}\log_2\left(\frac{1}{\bar{C}^2}\right)<\frac{1}{\bar{C}}\Rightarrow \frac{4}{\bar{C}}\log_2\bar{C}<1$ and that holds for $\bar{C}>16$. From equation (\ref{eqn:equality_for_inner_rate}) we get that $\bar{C}>16$ for the networks where our code construction would work or else the rate of the inner code would be negative.

So for the values of interest of $\bar{C}$ as long as we set $2P_I\leq^{\ref{eqn:probability_of_error_inner_code}} 82^{-\ell}\leq \frac{1}{\bar{C}^2}\Rightarrow \ell=3+\lceil\log_2\bar{C}\rceil$ we are certain that $h(2P_I)\leq \frac{1}{\bar{C}}$ for values of $\bar{C}$ that are of interest to us.

\section{Chernoff bound}
\label{appen:chernoff1}

For any random variable $A$ and for every $t\geq 0$ 
\begin{align*}
\Prob\left(A\geq a\right)= \Prob\left(tA\geq ta\right)= \Prob\left(e^{tA}\geq e^{ta}\right)\displaystyle\mathop{\leq}^{(*)} \frac{\stexp\left(e^{tA}\right)}{e^{ta}}
\end{align*}
where $(*)$ is the Markov inequality. If we assume that $A=\sum_{i=1}^{K}A_i$ where $A_i$ are independent identical distributed random variables with $\Prob(A_i=1)=q$ and $\Prob(A_i=1)=1-q$ then the above inequalities become
\begin{align*}
\Prob\left(\displaystyle\sum_{i=1}^K A_i\geq a\right)&\leq \frac{\stexp\left(e^{t\sum_{i=1}^{K}A_i}\right)}{e^{ta}}= \frac{\stexp\left(\displaystyle\prod_{i=1}^{K}e^{tA_i}\right)}{e^{ta}}=\\ &=\frac{\displaystyle\prod_{i=1}^{K}\stexp\left(e^{tA_i}\right)}{e^{ta}}= \frac{\left( q e^t+1-q\right)^K}{e^{ta}} 
\end{align*}
or since $t$ is chosen arbitrarily one can get the tightest bound by
\begin{align*}
\Prob\left(\displaystyle\sum_{i=1}^K A_i\geq a\right)\leq \min_{t>0}\frac{\left( q e^t+1-q\right)^K}{e^{ta}}. 
\end{align*}
The minimum value is attained for $t=\ln\left(\frac{a(1-q)}{q(n-a)}\right)$ and the minimum value gives the following bound
\begin{align}
\Prob\left(\displaystyle\sum_{i=1}^K A_i\geq a\right)\leq K^K\left(\frac{1-q}{K-a}\right)^{K-a}\left(\frac{q}{a}\right)^a
\label{eqn:app_upper_bound1}
\end{align}

For our case $a=2Kq$ and therefore equation $(\ref{eqn:app_upper_bound1})$ becomes
\begin{align*}
\Prob\left(\displaystyle\sum_{i=1}^K A_i\geq 2Kq\right)&\leq \left[\text{Exp}\left((1-2q)\ln\left(\frac{1-q}{1-2q}\right)- 2q\ln(2)\right)\right]^K 
\end{align*}
It's not difficult to show that
\begin{align*}
(1-2q)\ln\left(\frac{1-q}{1-2q}\right)- 2q\ln(2)\leq -0.3q
\end{align*}
for all $q<\frac{1}{2}$, therefore the above inequality becomes
\begin{align*}
\Prob\left(\displaystyle\sum_{i=1}^K A_i\geq 2Kq\right)\leq \text{Exp}\left(-0.3Kq\right)
\end{align*}

In our problem $K=N$ and $q$ is equal to the upper bound of $P_I$ that is equal to $4\hspace{0.5mm} 2^{-\ell}$. Therefore
\begin{align*}
\Prob\left(\text{More than twice the expected number of bit flips}\right)&\leq \Prob\left(\text{More than twice the expected number of symbol errors}\right)\leq\\ 
&\leq\text{Exp}(-0.3\frac{n}{\ell}4 2^{-\ell})\approx \text{Exp}(-0.15\frac{n}{\bar{C}\log_2 \bar{C}})  
\end{align*}

\end{appendices}
\end{document}